\newif\ifapj    
\apjfalse
\newif\ifpng	
\pngtrue

\ifapj
  \documentclass[apj]{emulateapj}
  \usepackage{float}
  \usepackage{natbib}
  \usepackage[normalem]{ulem}
\else
  \documentclass[usenatbib]{mn2e}
  \usepackage{graphicx}
  \usepackage{booktabs}
  \usepackage{threeparttable}
  \def\lesssim{\mathrel{\hbox{\rlap{\hbox{\lower5pt\hbox{$\sim$}}}\hbox{$<$}}}}
  \def\gtrsim{\mathrel{\hbox{\rlap{\hbox{\lower5pt\hbox{$\sim$}}}\hbox{$>$}}}}
\fi

\newcommand{\Ntot}{15}

\title[LAEs around QSO SDSSJ114514.18+394715.9]{Discovery of an Overdensity of Lyman-alpha Emitters Around a $\mathrm{z}\sim4$ QSO with the Large Binocular Telescope}
\author[Adams et al.]
	{\parbox{18cm}{Scott M. Adams$^{1}$, Paul Martini$^{1,2}$, Kevin V. Croxall$^{1}$, Roderik A. Overzier$^{3,4}$ and John D. Silverman$^{5}$}
\\
\\
$^{1}$ Dept.\ of Astronomy, The Ohio State University, 140 W.\ 18th   Ave., Columbus, OH 43210\\
$^{2}$ Center for Cosmology and AstroParticle Physics (CCAPP), The Ohio State University, 191 W.\ Woodruff Ave., Columbus, OH 43210\\
$^{3}$ Department of Astronomy, University of Texas at Austin, 1 University Station C1400, Austin, TX 78712, USA\\
$^{4}$ Observat\'orio Nacional, Rua Jos\'e Cristino, 77. CEP 20921-400, S\~ao Crist\'ov\~ao, Rio de Janeiro-RJ, Brazil\\
$^{5}$ Kavli Institute for the Physics and Mathematics of the Universe (Kavli IPMU, WPI), Todai Institutes for Advanced Study, \\
  the University of Tokyo, Kashiwa 277-8583, Japan\\
E-mail: sadams@astronomy.ohio-state.edu}
\begin{document}
\voffset -1.5cm
\maketitle

\begin{abstract}
Measurements of QSO clustering in the SDSS show that $\mathrm{z}>4$ QSOs are some of the most highly biased objects in the
Universe. Their large correlation lengths of $r_0 \sim 20h^{-1}$Mpc are
comparable to the most massive clusters of galaxies in the Universe today and
suggest that these QSOs may mark the locations of massive cluster
progenitors at high redshift.
We report the discovery of an overdensity of LBGs around QSO SDSSJ114514.18+394715.9 as part of our survey to  identify Lyman-Break galaxies (LBGs) around luminous $\mathrm{z}\sim4$ QSOs.
In this field three of the eight LBGs with secure redshifts are consistent with the redshift of the QSO.  
We find that the likelihood that this is merely an apparent overdensity due to the chance selection of field galaxies is only 0.02\%, based on comparisons to simulations and our modeled selection efficiency.
Overall, our survey finds four of the $\Ntot$ LBGs with secure redshifts are consistent with the redshifts of their respective QSOs, which is consistent with luminous QSOs residing in larger haloes.
\end{abstract}


\section{Introduction}

One of the major, missing links in the study of structure formation has been
the assembly of the most massive clusters of galaxies. Many observations of
cluster galaxies in the present-day Universe have shown that the most massive
cluster galaxies formed their stars earlier than comparably massive field
galaxies \citep[e.g.][]{Kelson97}.

Observations have shown that while star formation increases with redshift in clusters of galaxies (the so-called Butcher-Oemler Effect), the rates remain below those found in comparable field galaxies to about $\mathrm{z}=1.5$ \citep{Eisenhardt08}. 
This redshift range appears to mark the point where the star formation rate in clusters of galaxies, or their progenitors, begin to have higher star formation rates than comparable field galaxies \citep{Brodwin13}.

Stellar population models of local cluster galaxies predict that cluster galaxies should have more star formation than field galaxies at $z>2$, yet it has proven extraordinarily
difficult to identify clusters at these high redshifts
 \citep[see][for a review]{Chiang13}.
 Cluster progenitors have been difficult to
identify for two reasons: the galaxies in these clusters are extremely
faint ($i_{AB}>24-27$ mag) and these systems are extremely rare
(one per several square degrees). As a consequence, many tens of square
degrees need to be surveyed to find truly massive cluster progenitors.

Proto-clusters found by blind searches include one at $\mathrm{z}\sim5.3$ with $>4\times10^{11}M_{\odot}$ discovered by \cite{Capak11} in the 2-square degree Cosmological Evolution Survey (COSMOS) field and one at $\mathrm{z}\sim6$ found by \cite{Toshikawa12} in the Subaru Deep Field.

An alternative approach is to search around highly biased objects. This
has been accomplished with searches around high-redshift radio galaxies (HzRGs)
\citep[e.g.][]{Roettgering94,Hayashi12,Rigby14}.
One of the most extensively studied $\mathrm{z}<4$ proto-clusters is
associated with TN J1338-1942 at $\mathrm{z}=4.1$
\citep[e.g.][]{Venemans02,Miley04,Intema06}. 
\cite{Venemans07}, \cite{Hatch11}, and \cite{Wylezalek13} also show that HzRGs are often embedded in overdense structures of galaxies.
($2.0<\mathrm{z}<4.1$; $L_{500 MHz} > 10^{28.5}$ WHz$^{-1}$.
A major disadvantage of this approach is that 
not much is known about the host halo population of HzRGs
.  Although two-point correlation function analysis has shown that, at a median redshift of $\mathrm{z}\sim1$, they reside in the most strongly clustered halos \citep{Overzier03}, at higher redshifts there is limited information that they are very massive galaxies hosting massive black holes \citep[e.g.,][]{Drouart14}.
 
An alternate approach that is physically well-motivated and
well-calibrated is to identify cluster progenitors with observations of
the most luminous high-redshift ($\mathrm{z}>3$) QSOs (such as the those discovered in the SDSS). A major advantage of
this sample is that the measured clustering is extremely large
\citep[$r_0>20h^{-1}$Mpc;][]{Shen07} and in fact is comparable to the most
massive clusters in the Universe at the present day. Yet what makes this
sample ideal for this search is that the clustering strength and space density
of these QSOs strongly constrains the minimum halo mass of the QSO hosts
\citep[e.g.][]{Martini01}. At $\mathrm{z}\sim4$ the minimum halo mass corresponds
to $8 \times 10^{12} M_\odot$ \citep{Shen07} and therefore nearly all
halos above this mass at $z=4$ will evolve into the $M > 10^{14} M_\odot$
halos characteristic of clusters today.

An additional implication of the QSO clustering measurement at $z>4$ is that
their very low space density and very large clustering strength implies
both a high QSO duty cycle (nearly unity) and very small scatter ($<0.3$ dex)
between QSO luminosity and halo mass \citep{White08,Shankar10}.
While the strong clustering implies that essentially all QSOs will be
associated with cluster progenitors, the high inferred duty cycle implies
that essentially all cluster progenitors will be associated with a
QSO. This means that observations of $z=4$ QSO fields may provide a representative
sample of the progenitors of the most massive clusters. The very small
scatter between QSO luminosity and halo mass indicates that halo mass
should be a strong function of QSO luminosity and thus the best technique
is to target the most luminous QSOs.

A few studies have used QSOs to look for proto-clusters.  \cite{Priddey08} find an excess of sub-mm galaxies in the fields of three luminous $\mathrm{z}>5$ QSOs.  \cite{Kim09} observed five $\mathrm{z}\sim6$ QSO fields with the Advanced Camera for Surveys (ACS) on \emph{Hubble}, and find two are overdense, two are underdense, and one has have an average density of $i_{775}$-dropout galaxies, though \cite{Overzier09mnras} point out that the fields examined by \cite{Kim09} could have overdensities on scales larger than the narrow field of view of ACS.  More recently,  \cite{Banados13} detected no evidence of an overdensity of Lyman-$\alpha$ emitters (LAEs) around a $\mathrm{z}\sim5.7$ QSO and \cite{Simpson14} did not find an excess of bright galaxies around a $\mathrm{z}\sim7.1$ QSO.  However, \cite{Husband13} detected overdensities around $\mathrm{z}\sim5$ QSOs and \cite{Utsumi10} found evidence of a proto-cluster around a $\mathrm{z}\sim6.4$ QSO.  
Given that the results of these studies are mixed and that detections from these studies are not uniform, the strength of any correlation between QSOs and over-densities remains unclear, especially on larger scales.

In this work we target the fields of nine of the brightest $\mathrm{z}\sim4$ SDSS QSOs with the Large Binocular Telescope (LBT).  We identify candidate $\mathrm{z}\sim4$ LBGs with deep $g'$, $r'$, $i'$, and $z'$ photometry and spectroscopically follow-up a limited number of LBG candidates in four of the QSO fields.  We find evidence for an overdensity of LAEs around the luminous $z=4.044$ quasar SDSSJ 114514.18+394715.9.  While we have fewer redshifts for the other fields, those data at least do not rule out overdensities.  
We start in \S2 with a description of the data and observations.  In \S3 we present the spectroscopically measured redshifts and compare these to the results expected from the Millennium Simulation for fields with and without a proto-cluster.  In \S4 we discuss the significance of our results and present our conclusions.
In this paper, all magnitudes are given in the AB system.  We assume a $\Lambda$CDM cosmology with $H_{0}=71$ km s$^{-1}$Mpc$^{-1}$, $\Omega_{M}=0.3$, and $\Omega_{\Lambda}=0.7$.

\section{Observations and Data Reduction}
\subsection{QSO Target Selection and LBC Imaging}

Our sample selection began with the 269 QSOs at $4<\mathrm{z}<4.1$ from SDSS DR7 \citep{Schneider10}.  We then selected the 28 QSOs with $M_{i} < -28$, which are those in the top 10\% in absolute magnitude.  The nine QSOs that we observed were largely the most luminous ones from this subsample, although it was somewhat dependent on coordinates (see Table \ref{table:qsos} for a complete list of targets).  
Virial black hole mass estimates based on the scaling with C IV line width and luminosity from \cite{Vestergaard06} for our targeted QSOs range from $2\times10^{9}$ to $2\times10^{10} M_{\odot}$ \citep{Shen11}.
We obtained 27 300s dithered integrations in $g'$ and $z'$-band and 9 300s dithered integrations in $r'$ and $i'$-band of each QSO field between December 2010 and April 2012 with the Large Binocular Camera (LBC; which has a $\sim575$ arc-minute$^2$ field of view) on the LBT using binocular mode.

\begin{table}
\caption{Targeted QSO Fields}
\begin{tabular}{rlrr}
\hline
\hline
{ID} & {Original z$^{a}$} & {Revised z$^{b}$} & {$M_{i'}$} \\
\hline
SDSSJ 012700.69-004559.2 & 4.0816 & $4.097 \pm 0.002$ & -28.952 \\
SDSSJ 024447.79-081606.0 & 4.0678 & $4.048 \pm 0.004$ & -28.955 \\
SDSSJ 094932.26+033531.7 & 4.0497 & $4.103 \pm 0.004$ & -29.203 \\
SDSSJ 095723.14+231849.4 & 4.0268 & $4.030 \pm 0.002$ & -28.589 \\
SDSSJ 095937.11+131215.4 & 4.0560 & $4.078 \pm 0.003$ & -29.620 \\
SDSSJ 105705.37+191042.8 & 4.0971 & $4.136 \pm 0.003$ & -28.501 \\
SDSSJ 114514.18+394715.9 & 4.0610 & $4.044 \pm 0.002$ & -28.783 \\
SDSSJ 122000.83+254230.7 & 4.0343 & $4.049 \pm 0.002$ & -29.040 \\
SDSSJ 152245.19+024543.8 & 4.0896 & $4.082 \pm 0.004$ & -28.319 \\
\hline
\hline
\end{tabular}
$^a$Redshifts from SDSS DR7 quasar catalog \citep{Schneider10} \\
$^b$Redshifts from \cite{Hewett10} reprocessing of the SDSS DR7 quasar catalog
\label{table:qsos}
\end{table}

These data were bias-subtracted, flat-fielded, and, in the case of the $z'$-band data, fringe-corrected with standard {\sc iraf} tasks.  Astrometry and stacking were performed with {\sc scamp} \citep{Bertin06} (with SDSS-DR7 serving as the astrometric reference catalog) and {\sc swarp} \citep{Bertin02}.  Sources were identified and magnitudes were measured using {\sc SExtractor} \citep{Bertin96} in two-image mode, with the $r'$, $i'$, and $z'$ mosaics coadded to form the detection image.  Magnitude zeropoints were found by comparing our aperture-corrected magnitudes to PSF magnitudes of stars retrieved from the SDSS DR8 \citep{Aihara11} SkyServer.  The typical RMS in the zeropoint calibration was $\sim0.04$ mag for $g'$, $r'$, and $i'$ and $\sim0.08$ mag for $z'$.  Magnitudes were corrected for Galactic extinction based on \cite{Schlegel98}.  The typical limiting magnitude for S/N $>$ 3 in our stacked images was $\sim27.5$ mag for $g'$ and $\sim25.5$ mag for $r'$, $i'$, and $z'$.

\begin{figure}
  \ifpng
    \includegraphics[width=9.2cm, angle=0]{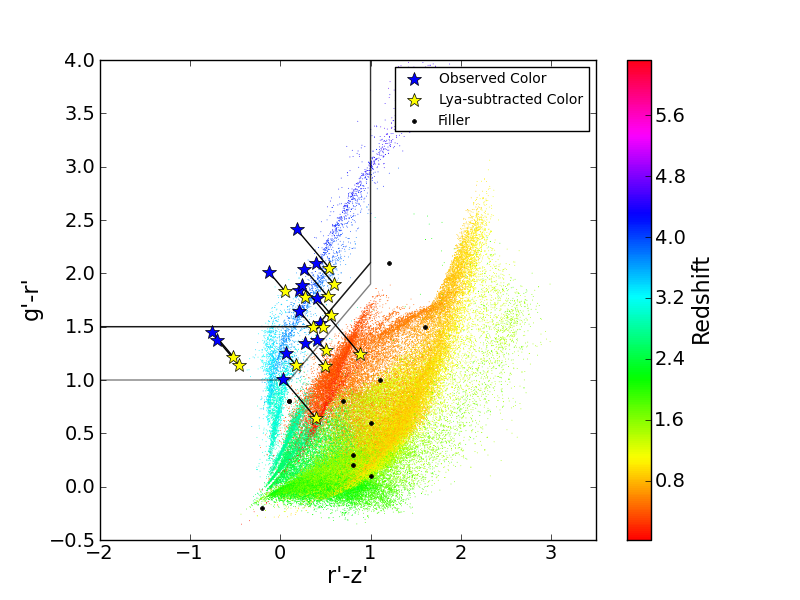}
  \else
    \includegraphics[width=9.2cm, angle=0]{fig1.eps}
  \fi
  \caption{Color-color plot of sources.  Sources from the simulated field catalog are shown as the small points, color-coded by their redshift.  The bounds of the color selection for the LBG4 sample is is shown by the thick black lines in the upper left corner.  The color selection for the LBG4E sample is the region between the thick and thin black lines at smaller $g'-r'$ than the LBG4 sample.  Targets from our LBG4 and LBG4E samples with spectroscopically measured redshifts are shown with large stars.  The large blue stars show the observed colors while the large yellow stars indicate the colors these targets would have without Ly$\alpha$ emission.  The colors of filler targets for which we spectroscopically determined redshifts are indicated by the small black points.  \label{fig:colorcolorplot}}
\end{figure}

\begin{figure}
  \ifpng
    \includegraphics[width=9.2cm, angle=0]{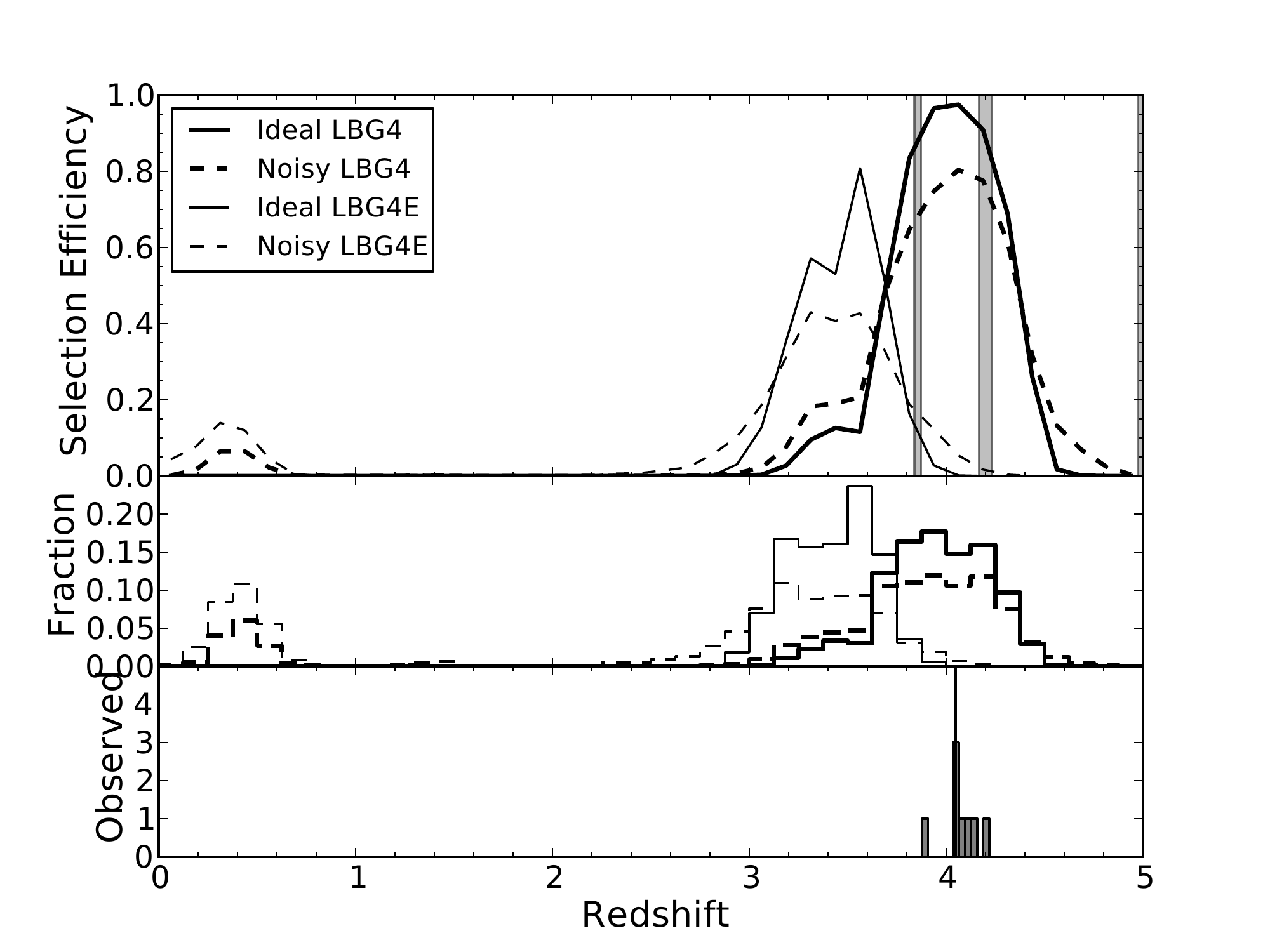}
  \else
    \includegraphics[width=9.2cm, angle=0]{fig2.eps}
  \fi
  \caption{The top panel shows the fraction of simulated ($z'<26$ mag) galaxies that meet our color selection as a function of redshift.  The shaded regions in the top panel are redshifts for which Lyman-alpha emission would appear at wavelengths affected by bright sky lines (causing the selection efficiency to drop below the nominal level).
The ``ideal" samples (solid lines) assume no photometric scatter.  Photometric scatter with a normal distribution with $\sigma=0.2$ mag has been added to the magnitudes in the ``noisy" samples (dashed lines).  These simulated galaxies do not include Ly$\alpha$ emission which would increase the selection efficiency of the LBG4E sample at $\mathrm{z}\sim4$.
The middle panel shows the relative numbers of galaxies expected to meet our color selection as a function of redshift.
The bottom panel, for comparison, shows a histogram of the redshifts of LBGs in the field of SDSSJ114514+394716.  The black vertical line in the bottom panel indicates the redshift of SDSSJ114514+394716. The Monte Carlo simulation we use to determine the odds of LBGs being within some $\Delta$z of the targeted quasar does fold in magnitude uncertainties, which results in a broader, less efficient selection function.
\label{fig:selection_efficiency}}
\end{figure}

We selected $\mathrm{z}\sim4$ LBGs (hereafter sample LBG4) based on the very successful technique employed by \cite{Yoshida06} in the Subaru Deep Field.  The criteria for LBG4 selection was $r'$ and $z'$ S/N $>$ 3, $g'-r' > 1.5$, $g'-r'>r'-z'+1.1$, $r'-z'< 1$, and $r'-z'>-2$ (see Fig. \ref{fig:colorcolorplot}).  This color selection is efficient for selecting LBGs with $3.8 \lesssim \mathrm{z} \lesssim 4.2$ (see Fig. \ref{fig:selection_efficiency} and \S\ref{sec:simulations}).  Unfortunately, our low S/N cut translates to large uncertainties in our observed colors.
While this cut includes more sources with lower probability of being within our target redshift range, the surface density of these sources was still sufficiently low that they could be assigned slits in our multi-object masks.

We also compiled a secondary sample of candidates (hereafter referred to as LBG4E) not included in the LBG4 sample by extending the LBG4 color cuts to $g'-r' > 1.0$ and $g'-r'>r'-z'+0.9$.  While the idealized selection efficiency of the LBG4E sample drops off by z$\sim4$, the magnitude measurement and zeropoint uncertainties effectively broaden the selection function so that it may include some z$\sim4$ LBGs that are missed by the LBG4 sample.

The observed overdensity factor expected to be associated with a proto-cluster is strongly tied to the redshift uncertainty of the selection technique.  As it can be very difficult to distinguish even massive cluster progenitors from random fields with color selection techniques \citep[see][Fig. 13]{Chiang13}, we also undertook a program of spectroscopic follow-up.

\begin{figure*}
  \ifpng
    \includegraphics[width=18.5cm, angle=0]{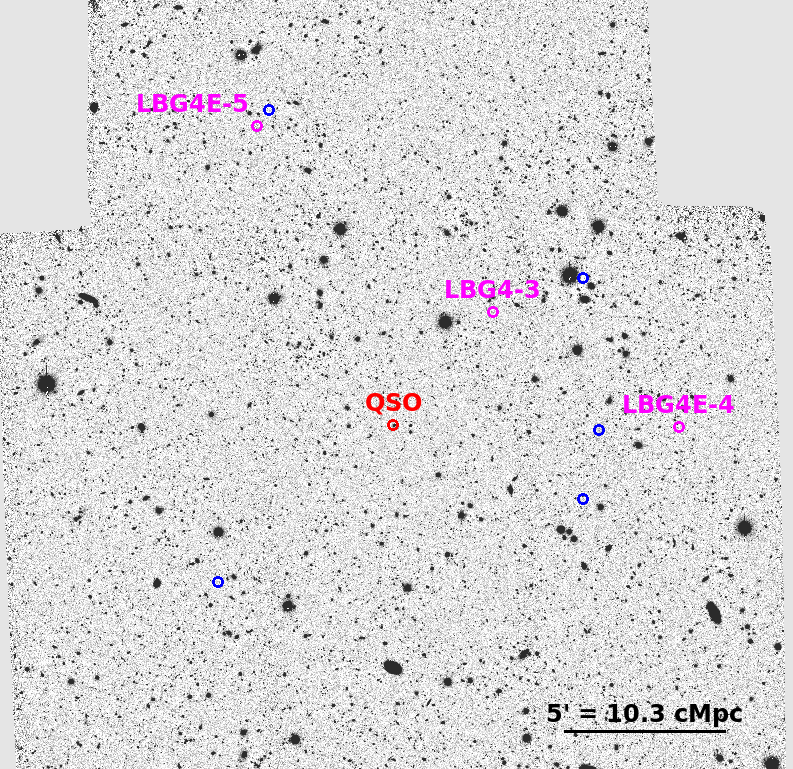}
  \else
    \includegraphics[width=18.5cm, angle=0]{fig3.eps}
  \fi
  \caption{LBC r' image centered on the quasar SDSSJ114514+394716 (shown in red).  This image was obtained by stacking 9 200s exposures.  In this field we successfully measured redshifts of eight LBGs.  The three labeled ones shown in magenta are consistent with the redshift of the quasar.  The five LBGs with redshifts inconsistent with that of the quasar are circled in blue.  The scale bar shows the comoving size at the quasar redshift of 4.044.\label{fig:qso1145imaging}} 
\end{figure*}

\subsection{MODS Spectroscopy}
We obtained low-resolution spectroscopy of 4 masks surrounding the QSO SDSSJ114514+394716 (see Fig. \ref{fig:qso1145imaging}) as well as 4 masks around 3 other QSOs with the first of the Multi-Object Double Spectrographs \citep[MODS1;][]{Pogge10} on LBT between November 2011 and June 2013 (see Table \ref{tab:masks}).  
The masks were selected based on target visibility with preference given to fields in which the LBG candidates appeared more clustered and/or had higher surface densities.
We used MODS1 in dual-prism mode with 1" slits, enabling wavelength coverage of $3300-10000\mathrm{\AA}$ with resolution of $R=500-150$ in a single exposure.  With the $6\times6$-arcminute field of view of MODS we were able to place slits on $\sim10$ LBG4 candidates in each mask.  As space allowed, we were able to place slits on $\sim10$ LBG4E candidates and $\sim30-35$ filler galaxies in each mask.  The LBG4 and LBG4E candidates ranged in magnitude from $24.3<r'<26.9$.

\begin{table}
\centering
\caption{Mask Exposure Times}
\begin{tabular}{lr}
\hline
\hline
{Mask Name} & {Exposures} \\
\hline
QSO0127m1 & 8x1800s \\
QSO1057m1 & 2x1800s \\
QSO1057m2 & 2x1800s \\
QSO1220m1 & 4x1800s \\
QSO1145m1 & 2x900s \\ 
QSO1145m2 & 2x900s \\ 
QSO1145m3 & 4x900s \\
QSO1145m4 & 3x900s \\
\hline
\hline
\end{tabular}
\label{tab:masks}
\end{table}

The MODS images were processed with MODS-specific bias-subtraction and flat-fielded procedures.  Cosmic rays were subtracted with {\sc l.a. cosmic} \citep{vanDokkum01}.  
We used a version of the publicly available {\sc xidl}\footnote{http://www.ucolick.org/$\sim$xavier/IDL/index.html}
 software library modified for MODS prism data to calculate the 2D wavelength solution and then sky-subtract, extract, and flux-calibrate the spectra.  
Flexure corrections were applied by shifting the wavelength solution by the offset of the measured centroid of the $6300\mathrm{\AA}$ sky line from its true wavelength (typically $\sim10\mathrm{\AA} \sim2\mathrm{pixels}$).

\section{Results}

\subsection{LBG Surface Density}

\begin{figure*}
  \ifpng
    \includegraphics[width=18cm, angle=0]{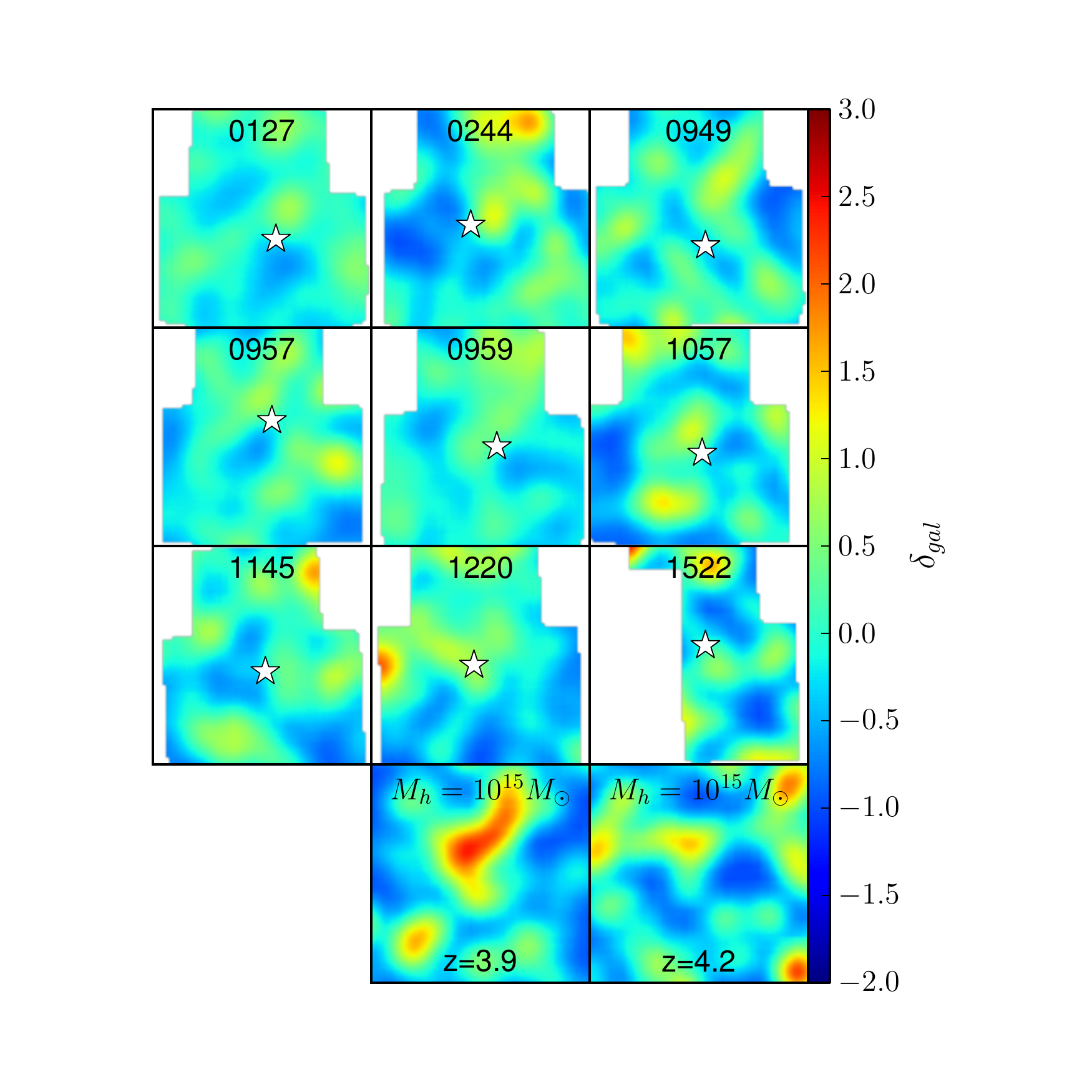}
  \else
    \includegraphics[width=18cm, angle=0]{fig4.eps}
  \fi
  \caption{Maps of surface density fluctuations, $\delta_{gal}=\frac{\Sigma-<\Sigma>}{<\Sigma>}$, of LBG4 sources in each of our targeted QSO fields (labeled with the first four digits of their RA) and two simulated snapshots of a $M_{halo}(\mathrm{z=0})=10^{15}M_{\odot}$ from the Millennium Run.  Each panel measures 20$\times$20$'$ with north up and east to the left.  The surface densities are smoothed with a Gaussian with FWHM $= 4'$ and $<\Sigma>$ is the surface density of LBG4 sources in the given field.  The LBG4 sources in the simulated fields are limited to r'$<$ 25 mag, which is comparable to the limiting magnitudes in the LBC fields.  There are no clear detections of overdensities of the photometrically selected LBG4 sources around the targeted QSOs.\label{fig-2D_maps}}
\end{figure*}

\begin{figure}
  \ifpng
    \includegraphics[width=9.2cm, angle=0]{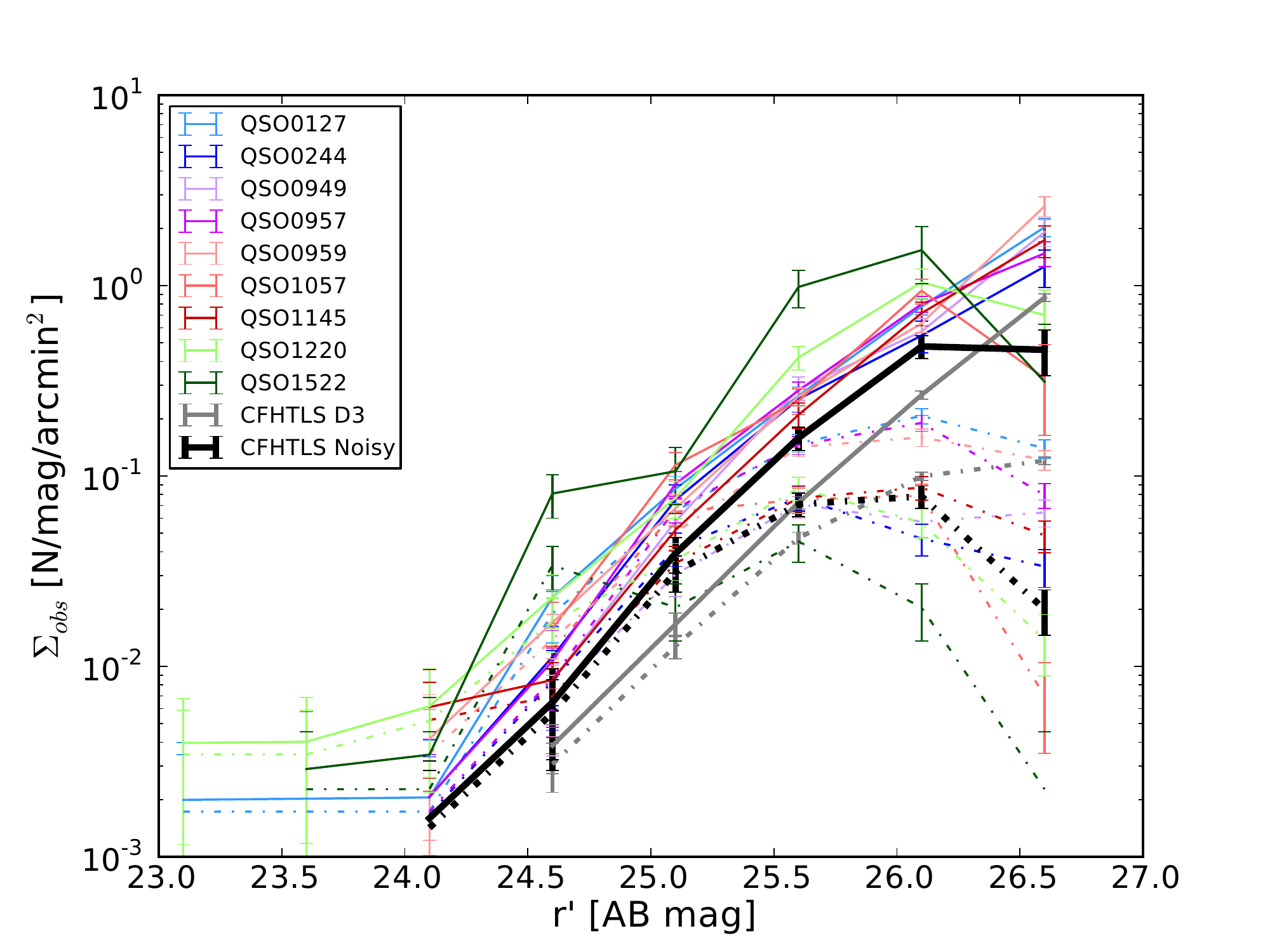}
  \else
    \includegraphics[width=9.2cm, angle=0]{fig5.eps}
  \fi
  \caption{Surface densities of galaxies satisfying our LBG4 selection criteria as a function of apparent r'-band magnitudes.  The dotted lines are the raw surface densities and the solid lines are the completeness-corrected surface densities.  The error bars are from Poisson statistics.  Although the LBG surface densities for our fields appear to be higher than that of the comparison CFHTLS deep field (CFHTLS D3 -- the thick gray line), the disparity is consistent with the differences in image quality.  The surface density of LBGs in the CFHTLS field after degrading the images to match the noise properties typical of our QSO fields (CFHTLS Noisy -- the thick black line) is not significantly different from that of our QSO fields.\label{fig-surface_density}}
\end{figure}

Our imaging data does not reveal clear overdensities of LBGs surrounding the targeted QSOs (see Fig. \ref{fig-2D_maps}).  We compare the surface densities of our samples of LBG4 galaxies in the QSOs fields with the density of sources that meet these selection criteria in the Canada--France--Hawaii Telescope Legacy Survey (CFHTLS) Deep 3 field after processing the CFHT images with the same pipeline (see Fig. \ref{fig-surface_density}.  We calculate completeness corrections as a function of magnitude by finding the fraction of injected sources that are recovered with our S/N thresholds.  After correcting for completeness our QSO fields have higher surface densities of LBGs than the CFHTLS field at all magnitudes.  However, it appears that this apparent overdensity is merely the result of our poorer image quality.  
We tested this by degenerating the seeing and image quality of the CFHTLS field to match the noise properties of a representative QSO field (QSO1145).  The completeness-corrected surface density of LBGs increased significantly with the increased photometric uncertainties as more interlopers scattered into our color-selection window.  The completeness-corrected surface density of LBGs in the degraded CFHTLS deep field is not significantly different from that of the QSO fields.

We also compare our surface densities with simulated galaxy catalogs from the Millennium Run Observatory (which will be described more fully in \S\ref{sec:simulations}) with our selection function.  Due to the large ($\Delta\mathrm{z}\sim0.5$) redshift uncertainty of our color selection, the limited number of galaxies expected to be detected with our imaging depth, and our limited field of view, it is possible for a real proto-cluster to escape detection.  A $\mathrm{z}=4.2$ snapshot of the progenitor of the most massive ($\mathrm{z}=0)$ halo in the Millennium Simulation does not have a significant overdensity at its center of $i'<25$ LBG4 galaxies relative to the average surface density of the 20$\times$20$'$ surrounding field, though an overdensity becomes more clear if the limiting magnitude is extended to $i'\sim27$ and the field is extended to 40$\times$40$'$.
With the limitations of our imaging data, we require spectroscopic follow-up to investigate the possible existence of overdensities associated with the QSOs.

\subsection{Redshifts}
\label{sec:redshifts}

\begin{table*}
\caption{Spectroscopically Measured LBGs}
\begin{tabular}{lllllccrrrrrr}
\hline
\hline
{ID} & {RA} & {Dec} & {Mask} & {z} & {Member$^{a}$} & {Interloper$^{b}$} & {g'} & {r'} & {i'} & {z'} & {EW$^{c}$ (\AA)} & {S/N$^{d}$} \\
\hline
LBG4-1 & 01:26:47.2 & $-$00:36:43 & QSO0127m1 & $4.107\pm0.003$ & Y & low 	   & 27.8 & 25.4 & 26.3 & 25.2 & $440\pm200$ & 29 \\	
LBG4E-1 & 01:26:53.8 & $-$00:35:54 & QSO0127m1 & $3.738\pm0.003$ & N & low 	   & 25.6 & 24.6 & 24.9 & 24.5 & $450\pm150$ & 44 \\	
LBG4E-2 & 10:56:24.3 & +19:10:35 & QSO1057m2 & $3.790\pm0.004$ & N & low 	   & 26.3 & 24.8 & 25.3 & 25.6 & $270\pm80$ & 20 \\	
LBG4E-3 & 11:44:43.8 & +39:44:58 & QSO1145m1 & $4.135\pm0.005$ & N & low  & 26.3 & 24.7 & 24.6 & 24.3 &  $40\pm15$ & 10 \\	
LBG4-2  & 11:44:41.3 & +39:47:06 & QSO1145m1 & $4.098\pm0.010$ & N & low        & 26.0 & 24.3 & 24.2 & 23.9 & $180\pm70$ & 12 \\	
LBG4E-4 & 11:44:28.4 & +39:47:11 & QSO1145m1 & $4.050\pm0.004$ & Y & low         & 26.3 & 24.9 & 24.7 & 24.5 & $180\pm100$ & 19 \\	
LBG4-3  & 11:44:58.2 & +39:50:45 & QSO1145m2 & $4.055\pm0.007$ & Y & low & 26.5 & 24.8 & 24.4 & 24.6 & $170\pm80$ & 13 \\	
LBG4-4  & 11:44:43.7 & +39:51:48 & QSO1145m2 & $3.908\pm0.004$ & N & low  	   & 27.4 & 25.4 & 25.2 & 25.5 & $200\pm100$ & 19 \\	
LBG4E-5 & 11:45:36.1 & +39:56:28 & QSO1145m3 & $4.042\pm0.009$ & Y & low 	   & 26.9 & 25.6 & 25.7 & 25.3 & $250\pm120$ & 11 \\	
LBG4-5  & 11:45:34.1 & +39:56:58 & QSO1145m3 & $4.067\pm0.009$ & N & low 	   & 27.0 & 25.2 & 24.9 & 25.0 &  $70\pm40$ & 10 \\ 
LBG4-6  & 11:45:42.2 & +39:42:26 & QSO1145m4 & $4.217\pm0.003$ & N & low     & 27.0 & 25.2 & 25.2 & 24.9 & $900\pm300$ & 32 \\	
LBG4-7  & 12:20:14.4 & +25:44:50 & QSO1220m1 & $4.320\pm0.003$ & N & low 	   & 26.2 & 24.2 & 23.9 & 23.9 & $300\pm15$ & 62 \\	
LBG4-8  & 12:20:07.2 & +25:43:17 & QSO1220m1 & $4.400\pm0.004$ & N & low 	   & 27.0 & 24.9 & 24.8 & 24.5 & $220\pm40$ & 28 \\	
LBG4E-6 & 12:20:12.4 & +25:44:43 & QSO1220m1 & $3.686\pm0.004$ & N & low 	   & 26.7 & 25.3 & 26.0 & 26.0 & $275\pm50$ & 35 \\	
LBG4E-7 & 12:19:54.6 & +25:45:51 & QSO1220m1 & $3.647\pm0.003$ & N & low 	   & 26.2 & 25.0 & 25.9 & 24.9 & $120\pm60$ & 17 \\	
\hline
\hline
\end{tabular}
\begin{flushleft}
$^a$Redshift is consistent with targeted QSO \\
$^b$Likelihood of a low-z interloper: low = no other emission lines visible; medium = a second emission line might be visible with low S/N; high = a second emission line is clearly detected \\
$^c$Observed-frame equivalent width of Ly$\alpha$ \\
$^d$S/N of Ly$\alpha$ emission
\end{flushleft}
\label{tab:redshifts}
\end{table*}

\begin{table*}
\caption{Spectroscopically Measured Filler Targets}
\begin{tabular}{lllllccrrrrr}
\hline
\hline
{ID} & {RA} & {Dec} & {Mask} & {z} & {Member$^{a}$} & {Interloper$^{b}$} & {g'} & {r'} & {i'} & {z'} & {EW$^{c}$ (\AA)} \\
\hline
FILL-1  & 01:26:46.2 & -00:38:58 & QSO0127m1 & $4.607\pm0.003$ & N & high & 22.2 & 22.0 & 21.6 & 21.2 & $135\pm20$ \\
FILL-2  & 01:27:03.3 & -00:36:13 & QSO0127m1 & $3.738\pm0.003$ & N & high & 22.6 & 21.8 & 21.4 & 21.1 & $25\pm8$  \\
FILL-3  & 01:26:50.7 & -00:35:02 & QSO0127m1 & $4.541\pm0.003$ & N & high & 22.5 & 22.2 & 21.9 & 21.4 & $60\pm40$  \\
FILL-4  & 01:26:54.1 & -00:34:47 & QSO0127m1 & $3.752\pm0.004$ & N & high & 22.9 & 21.9 & 21.2 & 20.8 & $55\pm20$  \\
FILL-5  & 10:56:26.3 & +19:11:36 & QSO1057m2 & $3.876\pm0.001$ & N & low         & 27.9 & 25.8 & 25.2 & 24.6 & $1000\pm900$ \\
FILL-6  & 10:56:31.2 & +19:10:02 & QSO1057m2 & $3.759\pm0.001$ & N & low        & 25.8 & 25.0 & 25.2 & 24.9 & $200\pm40$ \\
FILL-7  & 10:56:47.8 & +19:11:09 & QSO1057m2 & $3.927\pm0.001$ & N & high       & 24.3 & 23.5 & 23.8 & 23.4 & $475\pm200$ \\
FILL-8  & 10:56:40.5 & +19 12:13 & QSO1057m2 & $4.551\pm0.002$ & N & low & 25.7 & 25.9 & 25.2 & 26.1 & $80\pm40$  \\
FILL-9  & 11:45:29.6 & +39:42:53 & QSO1145m4 & $4.839\pm0.004$ & N & high & 24.9 & 24.8 & 24.2 & 23.8 & $80\pm40$  \\
FILL-10 & 11:45:48.1 & +39:42:59 & QSO1145m4 & $4.323\pm0.003$ & N & low & 25.7 & 25.1 & 24.5 & 24.1 & $120\pm70$ \\
FILL-11 & 12:20:03.3 & +25:46:03 & QSO1220m1 & $4.251\pm0.005$ & N & low       & 27.8 & 26.3 & 25.7 & 24.7 & $230\pm50$ \\
\hline
\hline
\end{tabular}
\begin{flushleft}
$^a$Redshift is consistent with targeted QSO \\
$^b$Likelihood of a low-z interloper: low = no other emission lines visible; medium = a second emission line might be visible with low S/N; high = a second emission line is clearly detected \\
$^c$Observed-frame equivalent width of Ly$\alpha$ emission
\end{flushleft}
\label{tab:filler}
\end{table*}

\begin{figure*}
  \begin{center}
    \ifpng
      \includegraphics[width=18.5cm, angle=0]{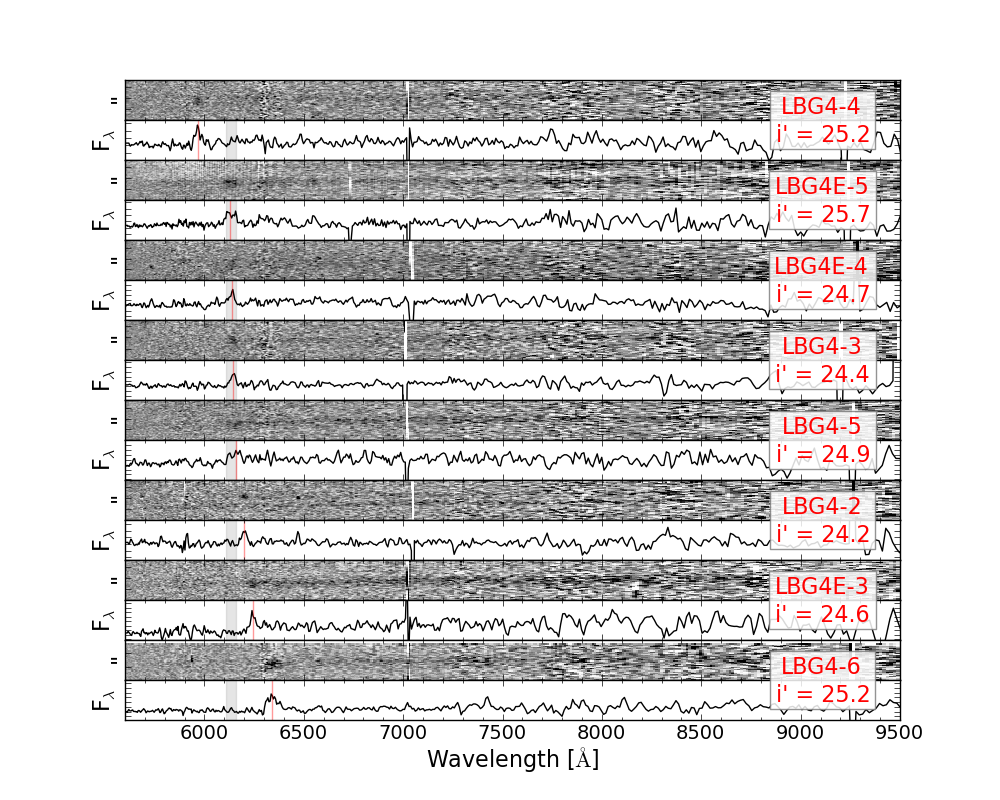}
    \else
      \includegraphics[width=18.5cm, angle=0]{fig6.eps}
    \fi
  \end{center}
  \caption{2D and 1D spectra of targets around SDSSJ114514.18+394715.9 for which we measured redshifts (see Table \ref{tab:redshifts}).
  The red vertical lines indicate the measured wavelength of Lyman-alpha emission for each object.  The shaded-gray regions indicate the wavelengths for which Lyman-alpha emission would be consistent with the redshift SDSSJ114514.18+394715.9 given our uncertainties ($\mathrm{z}=4.044\pm0.02$).  LBG4E-5, LBG4E-4, and LBG4-3 have emission lines within this window and are taken to be consistent with the redshift of the nearby quasar, while the other five spectra are inconsistent with the quasar redshift.\label{fig-2dspectra}}
\end{figure*}

\begin{figure*}
  \begin{center}
    \ifpng
      \includegraphics[width=18.5cm, angle=0]{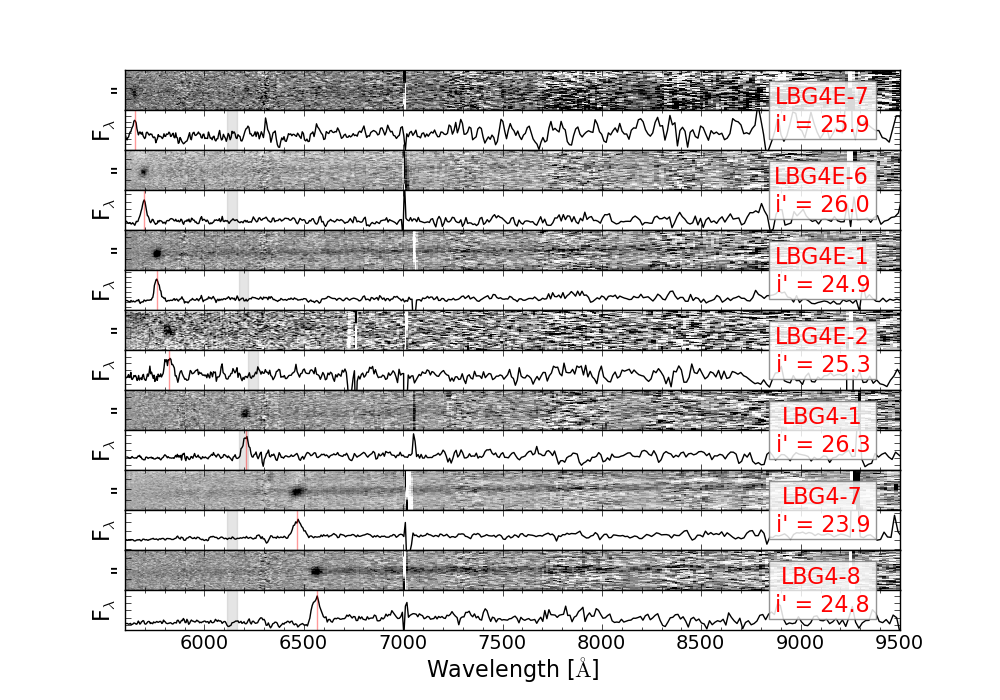}
    \else
      \includegraphics[width=18.5cm, angle=0]{fig7.eps}
    \fi
  \end{center}
  \caption{2D and 1D spectra of LBG4 and LBG4E targets for the other QSO fields for which we measured redshifts (see Table \ref{tab:redshifts}).
The red vertical lines indicate the measured wavelength of Lyman-alpha emission for each object. The shaded-gray regions indicate the wavelengths for which Lyman-alpha emission would be consistent with the targeted QSO given our uncertainties ($\mathrm{z} = 4:044 \pm0.02$).  LBG4-1 has an emission lines within this window and is taken to be consistent with the redshift of the nearby
quasar, while the other six spectra are inconsistent with their quasar redshifts.\label{fig-other2dspectra}}
\end{figure*}

The accurate identification of galaxies associated with the targeted quasars is very sensitive to the redshift measurement of the quasars.  However, quasar redshifts are notoriously difficult to measure due to the possible presence of significant quasar outflows and velocity shifts of different emission lines \citep{Gaskell82,Richards02}.  As described in \cite{Schneider10} the redshifts in the SDSS DR7 quasar catalog are determined by fits to template spectra.  \cite{Hewett10} find that the SDSS DR7 quasar catalog contains systematic biases of $\Delta z / (1+z) \geq 0.002$, and present revised redshift measurements for high-redshift quasars that include the cross-correlation of CIV.  In Table \ref{table:qsos} we separately list the redshifts for the quasars in our sample determined by both \cite{Schneider10} and \cite{Hewett10} to illustrate the systematic uncertainty in the redshift measurements, but for our analysis we only use the revised redshifts from \cite{Hewett10}.

We successfully measure redshifts (based on Ly$\alpha$ emission) of $\Ntot$ LBG4 and LBG4E galaxies, including eight from the masks around the SDSSJ114514.18+394715.9 QSO (see Figs \ref{fig-2dspectra} and \ref{fig-other2dspectra}).   This represents only a small fraction of the photometrically-identified LBG4 and LBG4E sources as $<10\%$ were followed up spectroscopically and only $\sim1/5$ of these had strong enough Ly$\alpha$ emission to make an accurate redshift determination. The redshift measurements are summarized in Table \ref{tab:redshifts} and \ref{tab:filler}.  The uncertainty in our LBG redshifts include uncertainty in the Ly$\alpha$ centroid ($\sim1-10\mathrm{\AA}$), the uncertainty in the centroid of the $6300\mathrm{\AA}$ sky line ($\sim1\mathrm{\AA}$)  used to calibrate the zeropoint offset of the wavelength solution, and uncertainty in the fit of the wavelength solution. Additionally, during the observations of the masks in the SDSSJ114514.18+394715.9 field the seeing ($0.5-0.6"$) was significantly better than the slit widths ($\sim1"$), so uncertainty also arises from the possibility that the targets were not perfectly centered in their slits.  Given these factors, we adopt a redshift uncertainty of 0.02 for all of our LBGs in the modeling of our observations described in \S\ref{sec:simulations}.  We find that four of the LBG redshifts are consistent with the redshifts of the targeted quasar, including three of the redshifts measured from masks around the SDSSJ114514.18+394715.9 QSO.  Though we do not treat the source as being associated with its QSO in our analysis, we note that an additional LBG around the SDSSJ114514.18+394715.9 QSO (LBG4-5) has a redshift that differs from the QSO by only slight more than our adopted redshift uncertainty (0.023 vs 0.02).

There is a possibility that a fraction of these targets could be low-redshift interlopers.  Dwarf stars should not contaminate our spectroscopic sample given that our redshift determination requires the detection of a strong emission line.  However, the presumed detection of Lyman-alpha could actually be an emission line with a longer rest-frame wavelength, such as the [OII] $3727\mathrm{\AA}$ line observed at $\mathrm{z}\sim0.7$.  Our photometric color selection should suppress the likelihood of this occurring, but with our photometric uncertainties, such a possibility cannot be excluded.  

The exact contamination rate of low-redshift galaxies is difficult to quantify as it is dependent on the S/N of the source (more low-redshift galaxies could scatter in with larger color uncertainties), the distribution of [OII] line equivalent of low-z galaxies (high [OII] equivalent widths needed for low-redshift galaxies to contaminate the LBG samples), and the magnitude of the source (the number ratio of low-redshift to LBGs is higher at brighter magnitudes).  Based on our simulated galaxy catalog (which will be described more fully in \S\ref{sec:simulations}) low-redshift galaxies would only produce a significant interloper fraction if they have observed-frame (rest-frame) [OII] equivalent widths greater than 300 (180) angstroms.  However, such large equivalent widths are likely extremely rare, since, for example, the largest (rest-frame) equivalent width measured in the HETDEX Pilot Survey of 284 $\mathrm{z}<0.56$ [OII]-emitting galaxies was $<70$ angstroms \citep{Ciardullo13}.  Without large [OII] equivalent widths low-z galaxies would need large ($>0.5$ mag) color measurement errors in order to contaminate the LBG samples.

Given the S/N of our spectra, we do not expect to observe any emission lines other than Lyman-alpha in a true LBG \citep[see, e.g.,][]{Shapley03}.
The estimated likelihood that each measured LBG is a low-redshift interloper is given in Table \ref{tab:redshifts}, with spectra where there is only one observed emission line given a ``low" likelihood and spectra with significant continuum emission blue-ward of the emission or a detection (with S/N $>$ 10) of a second emission line are assigned a ``high" likelihood.

In principle low-redshift interlopers could be differentiated from LAEs based on the emission line profiles \citep{Rhoads03,Kashikawa06}.  The emission line profile of Lyman-alpha from luminous, high-redshift galaxies tends to be asymmetric due to the presence of outflows, whereas the line profile from the less luminous, low-redshift interlopers should be unskewed \citep[e.g.,][]{Stern99,Pettini01,Tapken07}.  We compute the skewness, $S$, as described in \cite{Kashikawa06} and the median is 0.2 (with a median uncertainty of 0.6).  This suggests that many (if not most) of our detections are indeed Ly-alpha, but this test does not allow us to unambiguously separate [OII] and Ly$\alpha$ emission for individual objects.

\subsection{Simulations}
\label{sec:simulations}
We use the Millennium Run Observatory \citep{Overzier13} to estimate the selection efficiency of our photometric selection and to determine the expected observational signal of a $\mathrm{z}\sim4$ proto-cluster in data of our quality.  The Millennium Run Observatory computes mock galaxy catalogs and images using semi-analytic galaxy formation models based on the suite of Millennium Run dark matter simulations \citep{Springel05}.  We used two mock catalogs in which, by construction, the $\mathrm{z}\sim4$ progenitor of the most massive cluster in the Millennium Run Simulations appears at the center of the simulated field.  In addition, we also use a mock catalog along a random line of sight from \cite{Henriques12} for comparison.  All mock catalogs used are based on the semi-analytic galaxy model from \cite{Guo11}.  Galaxy magnitudes and colors were calculated using the \cite{Bruzual03} stellar synthesis library attenuated by dust as described in \cite{Henriques12}.  Attenuation of the rest-frame UV by neutral hydrogen absorption in the intergalactic medium was applied using the \cite{Madau95} prescriptions \citep[see][]{Overzier13}.  Although the assumed cosmology was WMAP1, the differences with respect to more recent cosmologies are known to be very small for the type of study performed here \citep{Chiang13}.  The semi-analytic galaxy catalogs are trimmed to only include galaxies with $z'<26$, which is well-matched to the depth of our LBC imaging.

We estimate the selection efficiencies of our LBG4 and LBG4E samples by finding the fraction of galaxies in a mock catalog of a 1x1-degree blank field that satisfy the color-cuts of the samples (see Fig.\ref{fig:selection_efficiency}).
We also use the mock catalog of the blank field to estimate the likelihood our observed concordance of LBG redshifts with the SDSSJ114514.18+394715.9 quasar could arise from only field galaxies.  We randomly draw from the mock galaxies that satisfy the estimated idealized selection function and add in a redshift error pulled from a normal distribution with $\sigma=0.02$, which roughly corresponds to the average uncertainty of our redshift measurements.  First, we consider the significance of finding three LBGs within 0.02 in redshift of the SDSSJ114514.18+394715.9 quasar by performing $10^{6}$ realizations of drawing eight galaxies.  The probability of finding three or more out of eight field galaxies within 0.02 in redshift of the SDSSJ114514.18+394715.9 quasar is $0.02\%$.  If one of the three galaxies is actually an interloper then the probability increases to $\sim1\%$.  This indicates that the overdensity of LBGs around SDSSJ114514.18+394715.9 has very high significance.  As expected, no other redshift range in our data has a significant overdensity.

While the significance of our detection is dependent on a simplified picture of our selection function, most effects not taken into account by our Monte Carlo simulation would decrease the likelihood that our observed signal could be due to the chance selection of field galaxies.  In particular, uncertainties in the colors of our candidates due to both Poisson noise in the candidate flux measurements (most LBG4 candidates have S/N $\sim4$) and the RMS of our zeropoint calibration ($\sim0.04$ mag for $g'$, $r'$, and $i'$ and $\sim0.08$ mag for $z'$) will scatter low and high-redshift interlopers into our spectroscopic follow-up samples.  Another consideration is that we were only able to confidently measure redshifts for candidates with strong Lyman-alpha emission, but the semi-analytic models used to generate the mock catalogs did not predict Lyman-alpha fluxes.  Subtracting off the flux of the Lyman-alpha emission (which falls within the $r'$ passband for $\mathrm{z}\sim4$ objects) from the candidates for which we successfully measured redshifts would decrease their $g-r'$ and increase their $r'-z'$ colors by 0.1-0.4 mag (see Fig. \ref{fig:colorcolorplot}).  This means that our selection for LAEs -- the candidates for which we are able to measure redshifts -- is effectively broader than the selection function we modeled with LBGs, and further increases the significance of the overdensity.  Night sky lines, however slightly decrease the significance of our result because they decrease our ability to measure certain redshifts (though, one of the Lyman-alpha lines we successfully detected did lie on top of the relatively bright [OI] $6300\mathrm{\AA}$ line), effectively narrowing the selection function (see Fig. \ref{fig:selection_efficiency}).

We compare our results with the expected observational signature of a proto-cluster.  We look at z$\sim3.9$ and z$\sim4.2$ snapshots of what evolves to be the most massive halo (M($\mathrm{z}=0$)$\sim10^{15}$M$_{\odot}$) in the Millennium Run.  Approximately 1/3 of $z'<26$ mag galaxies appearing within $10'$ of the proto-cluster center in these snapshots that satisfy our color selection are within 0.02 in redshift of the proto-cluster (compared to just $2\%$ of galaxies in a blank field).  Employing binomial statistics with this fraction, the likelihood of observing three or fewer out of eight galaxies within this redshift window if there is a massive proto-cluster in the field is 75\% and the likelihood of three or more is 51\%.  Since the selection efficiency of our color-cuts is slightly higher for $\mathrm{z}=4.044$ than for $\mathrm{z}\sim3.9$ or $\mathrm{z}\sim4.2$, the fraction of color-selected galaxies that we would expect to be consistent with a massive halo at that redshift is somewhat larger than 1/3, but even so, our observed concordance of LBG redshifts with the SDSSJ114514.18+394715.9 quasar is consistent with a massive halo.

We also consider the significance of our aggregate results for all QSO fields and for all QSO fields except SDSSJ114514.18+394715.9.
Again we draw galaxies from a mock catalog of a blank field with $10^{6}$ realizations.  Drawing the $\Ntot$ galaxies around the four quasar fields we observed, we find four or more are measured to be within 0.02 in redshift of their respective quasars only $0.02\%$ of the time.  If we subtract out the galaxies we detect around SDSSJ114514.18+394715.9, we are left with one out of seven LBGs within 0.02 in redshift of their respective quasars.  The Monte Carlo simulation shows that this situation could occur due to chance with blank fields $18\%$ of the time.  While our overall result shows an enhancement of LAEs around luminous quasars, the statistical significance of the result is due to the overdensity associated with SDSSJ114514.18+394715.9.


\section{Discussion and Conclusions}
We find a significant overdensity of LAEs around the luminous $\mathrm{z}=$ 4.044 quasar SDSSJ114514.18+394715.9.  Three of the eight color-selected galaxies with spectroscopic redshifts are within $\mathrm{z_{QSO}} \pm$ 0.02.  This level of overdensity is consistent with that expected from a massive, M($\mathrm{z}=0$)$\sim10^{15}$M$_{\odot}$ halo, but the small size of the sample of galaxies with measured redshifts means that the overdensity is also consistent with smaller halos.  Given the steepness of the halo mass function, there are many more $10^{14}$M$_{\odot}$ haloes than $10^{15}$M$_{\odot}$ haloes.
Consequently, it is probably more likely that our detections are the result of drawing more member LBGs from a lower mass halo by chance than of drawing the expected fraction from a $10^{15}$M$_{\odot}$ halo.
The spatial distribution of our candidate proto-cluster members in the SDSSJ114514.18+394715.9 field is significantly extended (see Fig. \ref{fig:qso1145imaging}), with one of the candidate members lying $\sim10$ cMpc (projected) from the quasar and the other two lying $\sim20$ cMpc away (as is the LAE with a redshift consistent with SDSSJ012700.69-0044559.2).  The confirmed nonmembers have a similar distance distribution.  \cite{Kashikawa07} found LAEs distributed around a $\mathrm{z}=4.87$ QSO, with a deficit within $\sim4.5$ Mpc.  \cite{Overzier09mnras} used simulations to show that typical $\mathrm{z}\sim6$ proto-clusters sizes do not exceed 25 cMpc, but note that the overdensities that proto-clusters sit in may extend beyond 30 arcmin in radius.  More recently, \cite{Chiang13} found that the characteristic size of large $\mathrm{z}\sim4$ proto-clusters is $13.0^{+3.8}_{-2.6}$ cMpc, though they also note that the overdensity associated with the proto-cluster could extend even farther.  For comparison, the projected size of the proto-cluster TN J1338-1942, the most massive $\mathrm{z}>4$ proto-cluster, is at least $2.7 \times 1.8$ Mpc \citep[$13.8 \times 9.2$ cMpc;][]{Venemans02}, although the fields of view of published images of the proto-cluster do not show its boundaries in all directions.  It is likely that not all of the LAEs that we identify as being associated with SDSSJ114514.18+394715.9 will fall into the halo, however, these LAEs would remain associated with the larger-scale structure around the massive, central halo.


Our aggregate result from all of our QSO fields is consistent with other recent work that suggests that luminous QSOs reside in high-mass haloes, but not necessarily in the highest-mass haloes.  \cite{Trainor12}, in a study of galaxy distributions around 15 of the most luminous $\mathrm{z}\sim2.7$ QSOs, found such QSOs inhabit haloes with mass $\mathrm{log}(M_{h}/M_{\odot})>12.1\pm0.5$ and comment that such haloes are more common by a factor of $\sim10^{6}-10^{7}$.  Simulations with a semi-analytic model presented in \cite{Fanidakis13} similarly show that while it is likely to find overdensities around the most luminous quasars, these enhancements are weaker than those expected for the most massive haloes.  Improved measurements of QSO clustering at these redshifts, and measurements of the luminosity dependence of QSO clustering, would also help to constrain the expected halo masses of these rare, highly biased objects.

Deeper, more extensive studies are necessary to 
determine more accurately the relation between QSO luminosity and halo mass.
Future surveys, such as the Subaru Hyper Suprime-Cam Survey will discover a large number of proto-clusters and should clarify the correlation between QSOs and proto-clusters.

\section*{Acknowledgements}
This paper uses data taken with
the MODS spectrographs built with funding from NSF grant
AST-9987045 and the NSF Telescope System Instrumentation
Program (TSIP), with additional funds from the Ohio Board of
Regents and the Ohio State University Office of Research. This
work was based on observations made with the Large
Binocular Telescope. The LBT is an international collaboration
among institutions in the United States, Italy, and Germany. The
LBT Corporation partners are: the University of Arizona on behalf
of the Arizona university system; the Istituto Nazionale di
Astrofisica, Italy; the LBT Beteiligungsgesellschaft, Germany,
representing the Max Planck Society, theAstrophysical Institute
Potsdam, and Heidelberg University; the Ohio State University;
and the Research Corporation, on behalf of the University of
Notre Dame, the University of Minnesota, and the University of
Virginia.  This paper made use of the modsIDL spectral data reduction pipeline developed in part with funds provided by NSF Grant AST-1108693.
The Millennium Simulation databases used in this paper and the web application providing online access to them were constructed as part of the activities of the German Astrophysical Virtual Observatory (GAVO).

\bibliography{../references}
\bibliographystyle{mn2e}

\end{document}